\documentstyle[12pt]{article}
\title{
SOLAR NEUTRINOS\\
WITH\\
THREE FLAVOR MIXINGS
}

\author{
	David Harley\(^1\) \and T.K. Kuo\(^2\) \and
James Pantaleone\(^1\)\thanks{
Address after August 1, 1992: Physics Department,
Indiana University,
Bloomington, IN  47405}\\
{ \it			\(^1\)Institute for Nuclear Theory}\\
{ \it			University of Washington}\\
{ \it			Seattle, Washington  98195}\\
				\and
{ \it			\(^2\)Physics Department}\\
{ \it			Purdue University}\\
{ \it			West Lafayette, IN  47907}\\
}

\date{}
\begin{document}

\begin{titlepage}

\maketitle

\begin{abstract}
The recent \(^{71}\)Ga solar neutrino observation
is combined with the \(^{37}\)Cl and Kamiokande-II
observations in an analysis for neutrino
masses and mixings.
The allowed parameter region is found for
matter enhanced mixings among all
three neutrino flavors.
Distortions of the solar neutrino spectrum
unique to three flavors are
possible and may be observed in
continuing and next generation experiments.
\end{abstract}

\end{titlepage}

For two decades the measurement of solar neutrinos
in the \(^{37}\)Cl experiment \cite{Cl} has found far fewer
neutrinos than predicted by the Standard Solar Model
(SSM).  This result was supported by the water Cherenkov
experiment Kamiokande-II \cite{KII} and, very recently,
by the gallium experiments SAGE \cite{SAGE} and GALLEX \cite{GALLEX}.
Theoretically, a very appealing explanation is that
resonant neutrino oscillations (the MSW effect \cite{MS,W}, for reviews
see \cite{KP,MSr}) can deplete the \(\nu_e\) flux produced in the sun.
The most interesting consequence of this interpretation is that
the observed neutrino flux is tied in with the neutrino
masses and mixing angles.
We have thus available the means to probe
neutrino masses as small as \(10^{-6}\) eV.
It should be noted that
such small neutrino masses are expected in the standard model
from effective gravitational corrections,
or they may come from new physics such as Grand Unified Theories \cite{AK}.
Thus solar neutrino observations provide an important window
into physics at very high energy scales.

However it is premature to draw definitive conclusions as to
the presence or absence of neutrino masses.
It is conceivable that the SSM may be incomplete.
Also, the present solar neutrino observations
generally suffer from low counting rates.
New solar neutrino detectors are needed to overcome these problems.
To aid the design choices for future experiments, we
here examine what neutrino mass and mixing parameters the present
data imply.
Indeed, such an analysis has been carried out \cite{GALLEX}
assuming only two flavors of neutrinos.
The result is that the neutrino masses and mixing angles are constrained
to lie in a very small region, almost pinpointing the neutrino
parameters.  This is an extremely important conclusion and one needs
to assess its reliability.  An obvious question is how good the two-flavor
approximation is.  The purpose of this Article is to analyze the problem
in the realistic case of three flavors.
We also discuss how the favored parameter regions
may be probed in the future.

The preferred neutrino mass and mixing parameter regions
are found by calculating the chi-squared between the
predicted and measured results.
The \(^{37}\)Cl and GALLEX results given in Table 1 are used.
The SAGE result is not used, pending further analysis of recent data.
The Kamiokande-II measurements in 12 different
energy bins are included explicitly
and their overall systematic error is correctly accounted for \cite{KII}.
The SSM fluxes of Bahcall and Pinsonneault \cite{SSM}
are used.  Their estimated uncertainties
are neglected herein since they are small compared to the
experimental uncertainties--including them would slightly increase
the allowed neutrino parameter region.

When a neutrino propagates through matter,
a resonance can occur between the vacuum neutrino mass
and the induced mass from the electron background.
For three flavors,
there are two two-flavor type resonances which can occur \cite{KP3}:
a resonance between the \(\nu_e\) and the \(\nu_3\) vacuum mass eigenstate and
a resonance between the \(\nu_e\) and the \(\nu_2\) vacuum mass eigenstate.
For each resonance there is a range of energies for
which the neutrino flux is rotated into a different flavor
that present experiments can not observe.
This energy range spans from
a sharp, lower (adiabatic) threshold to a
more gradual, upper (nonadiabatic) threshold.
To leading order in small mixing, these energy ranges can be
qualitatively described as
\begin{equation}
6 MeV \left( {m_i^2 \over {10^{-4} eV^2}} \right) < E <
10 MeV \left({ {m_i^2 \times |U_{ei}|^2} \over {10^{-8} eV^2}} \right)
\end{equation}
for the  \(\nu_e-\nu_i\) resonance.
Here \(m_i\) is the ith neutrino's mass in vacuum, and
\(|U_{ei}|^2\) is the amount of \(\nu_e\) in the ith vacuum mass eigenstate.
This equation will aid in understanding the figures,
however for the actual calculations plotted, a more accurate,
analytical expression is used
to describe the survival probability \cite{KP3} of a solar \(\nu_e\).
That expression is valid for both small and large mixing,
and has the proper level crossing
behaviour in the convective zone.
The possibility of enhanced
mixing from propagation through the Earth is ignored--this
distorts the allowed parameter region slightly.

In certain limits, one neutrino decouples and then
the approximation of assuming only two flavors is reasonable.
This is the case at the top of Figs. (1) and (2).
There the plotted allowed
regions are in agreement with those found in ref. \cite{GALLEX}.
Generally, the region at small mixing corresponds to
resonant conversion of some of the \(^7\)Be and predominantly the
lower energy \(^8\)B solar fluxes.
For the region at large mixing,
resonant conversion occurs for all of the \(^8\)B and
varying amounts of the \(^7\)Be solar fluxes--but because of the large vacuum
mixing the observability of these fluxes is increased and that
of the pp flux is decreased.
These two two-flavor solutions can be easily distinguished
by future experiments.
In SNO \cite{SNO} or Super-Kamiokande \cite{SK},
the small mixing solution produces a "nonadiabatic"
distortion of the \(^8\)B flux
while the large mixing solution produces day-night differences \cite{dn}.
In BOREXINO \cite{BOREXINO},
we find that the small and large mixing solutions
predict 0.2-0.8 and 0.45-0.75 of the \(^7\)Be SSM flux, respectively.
However these two-flavor conclusions can be
modified by three-flavor effects.

One way that three-flavors can be relevant is if parts of both of
the two two-flavor conversion  energy ranges lie in the
experimentally observable region.
This is illustrated in Fig. (1).
At the top of the figure, the 3 neutrino decouples
because the \(|U_{e3}|^2\) mixing is small and also
\(m_3^2\) is large enough so that the conversion energy range of
the \(\nu_e-\nu_3\) resonance, Eq. (1) with i=3,
lies above that experimentally accessible with solar neutrinos.
But as \(m_3^2\) approaches \(1 \times 10^{-4}\) eV\(^2\)
the \(\nu_e-\nu_3\) resonance rotates away more of the higher energy
neutrinos.  For the region at small \(|U_{e2}|^2\),
this can be compensated for by decreasing
\(|U_{e2}|^2\) which decreases
the energy range where \(\nu_e-\nu_2\) resonant conversion occurs.
Since a single, two-flavor, adiabatic threshold in the \(^8\)B
flux can not explain the K-II data, \(|U_{e2}|^2\) can not
become too small.  However  the two-flavor range of
\(|U_{e2}|^2\), \(9\times 10^{-4}\) to \(3\times 10^{-3}\),
now extends down to \(1\times 10^{-4}\),
producing the "foot"
visible on the region at small \(|U_{e2}|^2\) \cite{other}.

As \(m_3^2\) decreases further, too much
\(^8\)B flux is rotated away and no solutions are possible.
But at \(m_3^2 \approx 5 \times 10^{-6}\) eV\(^2\)
the \(\nu_e-\nu_3\) resonance
explains the data with the \(\nu_e-\nu_2\) resonance decoupled.
Hence the lower allowed region extends to arbitrarily small \(|U_{e2}|^2\)
and \(m_2^2\) values.
This region terminates at
\(m_2^2 \times |U_{e2}|^2 \approx 2\times 10^{-10}\) eV\(^2\)
where the nonadiabatic threshold of the
\(\nu_e-\nu_2\) resonance converts too much of the
low energy solar pp flux.
\(m_3^2\) can increase near this border
because rotating away more of
the lower energy pp flux with the \(\nu_e-\nu_2\)
resonance can be compensated for
by removing less of the \(^7\)Be flux with the \(\nu_e-\nu_3\) resonance.
BOREXINO would then observe the full SSM \(^7\)Be flux.

Fig. (1) illustrates a general theme for three flavor effects.
Parts from both of the two two-flavor energy ranges
(also vacuum oscillation distortions \cite{justso})
can lie in the experimentally observable region at the same time.
The resulting spectral distortions are different from those
of the two-flavor case and, depending on the values of the
parameter, can be very complicated.
New experiments will be able to search for these spectral distortions.
Small distortions of the high energy \(^8\)B flux will be observable in SNO
and Super-Kamiokande.
Discerning distortions of the low energy pp neutrinos
may be possible with high statistics in the \(^{71}\)Ga experiments
in combination with accurate measurements of the other fluxes.
Additionally, expanding the energy range of observations would
expand the constraints on the "third" neutrino.
This may be done by observing the hep neutrinos (SNO and Super-Kamiokande)
and by using new techniques to observe the large
flux of low energy pp neutrinos (e.g. \cite{He}).

Another way that all three flavors can be relevant to the solar
neutrino flux is through the mixing.
Precision measurements at an energy threshold of a single resonance
can show the presence of a "third"
neutrino if the mixing with it is nonzero
(see Sect II.D.4 of ref. \cite{KP}).
Large mixings must be considered a likely possibility
since neutrino masses are obviously very different from
the known Dirac masses.
The effects of large mixings are illustrated in Fig. (2).

In Fig. (2), \(m_3^2\) is large enough so that the
energy range where \(\nu_e-\nu_3\) resonance conversion occurs
is above that relevant to current solar neutrino experiments.
However this "third" neutrino influences the observations through
the vacuum mixing.
At the top of Fig. (2), the mixing with the 3 neutrino is
small enough that the \(\nu_e-\nu_3\) resonance decouples.
However with increasing \(|U_{e3}|^2\) the vacuum mixing
rotates away more of the solar neutrinos.
Then three qualitatively new allowed regions appear.
On the right, the allowed region gradually grows an extension
along \(m_2^2 \approx 10^{-4}\) eV\(^2\) where
only the higher energy \(^8\)B neutrinos are resonantly rotated away.
On the left,  there abruptly appears a thin wall of an allowed region
along \(m_2^2 \times |U_{e2}|^2 \approx 6 \times 10^{-9}\) eV\(^2\).
This corresponds to resonant conversion of only the lower energy
\(^8\)B neutrinos.  For both of these new, allowed regions
the \(^7\)Be and pp neutrino fluxes are reduced by
the large, three-flavor, vacuum mixing.
In addition, Fig. (2) shows
a large allowed region at \(|U_{e3}|^2 \approx 0.5\).
Generally, this corresponds to
partial resonant conversion of all of \(^8\)B,
varying amounts of the \(^7\)Be, and 0.5 vacuum
mixing of the pp solar fluxes.

Fig. (2) is partially motivated by recent
measurements of the atmospheric neutrino flux \cite{atmos}.
One explanation of those observations is with \(m_3^2 = 1.0 \times 10^{-2}\)
eV\(^2\) and large \(|U_{e3}|^2\).
That parameter range can be further probed through long baseline
experiments using accelerator neutrinos \cite{LB}.
However note that results similar to Fig. (2) are
obtained in a plot of \(m_3^2, |U_{e3}|^2\), and \(|U_{e2}|^2\),
while keeping \(m_2^2\) "small",
\(10^{-11}\) eV\(^2\) \( < m_2^2 < 10^{-8}\) eV\(^2\).
This parameter region can be probed
by looking for seasonal variations in the solar \(^7\)Be line flux
\cite{PP,justso}.
BOREXINO will observe this flux at a high counting rate,
10-50 events/day, so that even very small seasonal variations
may be discernable.

There are generally two types of genuine
three-flavor effects.
If \(10^{-9}\) eV\(^2 < m_2^2 <  m_3^2 < 10^{-4}\) eV\(^2\),
the solar neutrino undergoes two resonances on its way out of the Sun.
If one \(|U_{ei}|^2 \geq 0.05\), then vacuum mixing effects
become important.  We have illustrated these two possibilities
by fixing one of the four parameters so that we may make
3-D plots of the allowed values of the neutrino parameters.
Other choices yield differences in detail, and
sometimes qualitatively new solutions.  We find that
assuming only two neutrino flavors is a simplifying
assumption that may easily be inadequate,
especially when accurate data become available.
When three flavors are included,
predictions for future experiments become much broader.
For BOREXINO, the \(^7\)Be flux can take any value between
0.2 to 1 of the SSM, and large seasonal variations are possible.
For SNO or Super-Kamiokande,
the mean \(^8\)B flux is already determined by the KII measurements,
but spectral distortions far more intricate than in the
two-flavor analysis are possible.

We would like to thank Sandip Pakvasa and Archie Hendry
for useful communications.
JP is grateful to the theory group at Brookhaven Laboratory
for their hospitality and partial support during the completion
of this work.

\raggedbottom
\pagebreak

\raggedbottom
\pagebreak

Table 1.  Results of solar neutrino experiments.  The flux is
given as a fraction of the SSM \cite{SSM} prediction.
\\

\begin{tabular}{| l | l | l | l |} \hline
Experiment & Process & E\(_{threshold}\) & Expt./SSM \\ \hline
Davis et al. & \(\nu_e + ^{37}\)Cl\( \rightarrow e + ^{37}\)Ar
   & 0.81 MeV  & 0.27 \(\pm\) 0.04 \\
Kamiokande-II & \(\nu + e \rightarrow \nu + e \)
   & 7.5 MeV  & 0.46 \(\pm\) 0.05 \(\pm\) 0.06 \\
SAGE & \(\nu_e + ^{71}\)Ga\( \rightarrow e + ^{71}\)Ge
   & 0.24 MeV  & 0.15 \(\pm\) 0.14 \(\pm\) 0.24 \\
GALLEX & \(\nu_e + ^{71}\)Ga\( \rightarrow e + ^{71}\)Ge
   & 0.24 MeV  & 0.63 \(\pm\) 0.14 \(\pm\) 0.06 \\ \hline
\end{tabular}
\raggedbottom
\pagebreak

\begin{center}
Figure Captions
\end{center}

The red surface \cite{graph} surrounds the region
allowed at 90\% confidence level
(for three variables)
by the solar neutrino observations.
The lines thereon correspond to the axes tic marks.
The "bottom" coordinate system is analogous to the
usual two-flavor type plot for the \(\nu_e-\nu_i\) resonance with
\(m^i_2\) (out of page) versus the mixing element
\(|U_{ei}|^2\) (horizontal axis).

\begin{enumerate}

\item UPPER FIGURE. The vertical axis is \(m_3^2\).  The constraints
\(|U_{e3}|^2\) = \(2 \times 10^{-3}\),
\(10^{-3} eV^2 > m_3^2 > m_2^2 > 10^{-8} eV^2\), and
\( 10^{-5} < |U_{e2}|^2 < 0.5\) are assumed.

\item LOWER FIGURE.
The vertical axis is \(|U_{e3}|^2\).  The constraints
\(m_3^2 = 1 \times 10^{-2}\) eV\(^2\),
\(|U_{e2}|^2 < 0.5\), and
\(10^{-2} < |U_{e3}|^2 < 0.5\)  are assumed.

\end{enumerate}

\raggedbottom
\pagebreak


\begin{thebibliography}{99}

\bibitem{Cl}
R. Davis, D.S. Harmer and K.C. Hoffman, Phys. Rev. Lett. 20, 1205 (1968);
R. Davis, in Proc. of the 21st Int. Cosmic Ray Conf., ed. by R.J. Protheroe
(University of Adelaide Press) 143 (1990).

\bibitem{KII}
K. Hirata et al., Phys. Rev. Lett. 65, 1297 (1990); 1301 (1990);
Phys. Rev. D44, 2241 (1991).

\bibitem{SAGE}
A.I. Abazov et al. (SAGE), Phys. Rev. Lett. 67, 3332 (1991).
T. Bowles, talk presented at NEUTRINO'92, Granada, Spain.

\bibitem{GALLEX}
P. Anselmann et al. (GALLEX), Phys. Lett. B285, 376 (1992);
390 (1992).

\bibitem{MS}
S.P. Mikheyev and A. Yu. Smirnov, Yad. Fiz. 42, 1441 (1985).

\bibitem{W}
L. Wolfenstein, Phys. Rev. D17, 2369 (1978); D20, 2634 (1979).

\bibitem{KP}
T.K. Kuo and J. Pantaleone, Rev. Mod. Phys. 61, 937 (1988).

\bibitem{MSr}
S.P. Mikheyev and A.Yu. Smirnov, Usp. Fiz. Nauk 153, 3 (1987).

\bibitem{AK}
S. Weinberg, Phys. Rev. Lett. 43, 1566 (1979).

\bibitem{SSM}
J.N. Bahcall and M.H. Pinsonneault, IASSNS-AST 92/10;
J.N. Bahcall and R.K.Ulrich, Rev.Mod. Phys. 60, 297 (1988).

\bibitem{KP3}
T.K. Kuo and J. Pantaleone,
Phys. Rev. Lett. 57, 1805 (1986);
Phys. Rev. D35, 3432 (1987).

\bibitem{MS3}
S.P. Mikheev and A.Yu. Smirnov, Phys. Lett. B200, 560 (1988).

\bibitem{SNO}
H.H. Chen, Phys. Rev. Lett. 55, 1534 (1985).
G.T. Ewan et al., Sudbury Neutrino Observatory (SNO) Proposal,
October (1987).

\bibitem{SK}
Y. Totsuka (Super-Kamiokande), Tokyo Univ. preprint ICRR-227-90-20.

\bibitem{dn}
K.S. Hirata et al., Phys. Rev. Lett. 66, 9 (1991).  See also
R.S. Raghavan et al., Phys. Rev. D44, 3786 (1991).

\bibitem{BOREXINO}
C. Arpesella, et al., BOREXINO at Gran Sasso, August (1991).

\bibitem{other}
The two-flavor, "large angle" solutions is not substantially modified
in Fig. (1), but it is modified for other values of
\(|U_{e3}|^2\).  For example, if \(|U_{e3}|^2 = 3\times10^{-4}\)
and \(m_3^2 \approx 10^{-5}\) eV\(^2\) then the allowed
range of \(m_2^2\) becomes \(10^{-5}\) to \(10^{-8}\) eV\(^2\)
(Harley, Kuo and Pantaleone, to be published).

\bibitem{justso}
J. Pantaleone, Phys. Rev. D43, R641 (1991).

\bibitem{He}
R.E. Lanou, H.J. Maris, and G.M. Seidel,
Phys. Rev. Lett. 58, 2498 (1987).

\bibitem{atmos}
E. Beier, et al., Phys. Lett. B283, 446 (1992);

\bibitem{LB}
J. Pantaleone, INT preprint DOE-ER-40561-059 (1992);
Phys. Lett. B246, 245 (1990);
R.H. Bernstein and S.J. Parke, Phys. Rev. D44, 2069 (1991).

\bibitem{PP}
S. Pakvasa and J. Pantaleone,
Phys. Rev. Lett. 65, 2479 (1990).

\bibitem{graph}
M. Lounsbery et al., SPIE Proceedings Vol. 1251 (1990) 94;
D. Meyers, S. Skinner and K. Sloan, Ass. Comp. Mach. Trans. Graph.,
Vol. 11, No. 3 (1992).

\end{thebibliography}
\end{document}